\documentclass[leqno,12pt]{amsart}
\usepackage{amssymb,amsmath}
\usepackage{amsthm}

\newcommand{\R}{{\mathbb R}}
\newcommand{\Z}{{\mathbb Z}}
\newcommand{\N}{{\mathbb N}}
\newcommand{\Q}{{\mathbb Q}}
\newcommand{\CA}{{\mathcal A}}

\newcommand{\CC}{{\mathcal C}}
\newcommand{\CE}{{\mathcal E}}
\newcommand{\CF}{{\mathcal F}}
\newcommand{\CH}{{\mathcal H}}

\newcommand{\CL}{{\mathcal L}}
\newcommand{\CM}{{\mathcal M}}
\newcommand{\CP}{{\mathcal P}}
\newcommand{\CS}{{\mathcal S}}
\newcommand{\CV}{{\mathcal V}}
\newcommand{\la}{{\langle}}
\newcommand{\ra}{{\rangle}}

\newcommand{\lin}{\operatorname{lin}}

\newcommand{\vol}{\operatorname{vol}}
\renewcommand{\a}{{\mathfrak{a}}}

\renewcommand{\c}{{\mathfrak{c}}}
\renewcommand{\d}{{\mathfrak{d}}}
\newcommand{\f}{{\mathfrak{f}}}
\renewcommand{\t}{{\mathfrak{t}}}
\newcommand{\p}{{\mathfrak{p}}}
\newcommand{\q}{{\mathfrak{q}}}
\newcommand{\lattice}{\Lambda}

\title{Summing   a polynomial function   over integral  points of a
polygon.  User's guide.}
\author{Velleda Baldoni}
\address{Universita  di Roma Tor Vergata,
Dipartimento di Matematica, via della Ricerca Scientifica, 00133
Roma, Italy} \email{baldoni@mat.uniroma2.it}

\author{Nicole Berline}
\address{Ecole Polytechnique, Centre de math\'ematiques Laurent Schwartz,
91128, Palaiseau, France} \email{berline@math.polytechnique.fr}

\author{Mich{\`e}le Vergne}
\address{Institut de Math\'ematiques de Jussieu, Th{\'e}orie des
  Groupes, Case 7012, 2 Place Jussieu, 75251 Paris Cedex 05, France}
\address{Ecole Polytechnique, Centre de math\'ematiques Laurent
  Schwartz, 91128, Palaiseau, France}
\email{vergne@math.polytechnique.fr}

\date{May 2009}

\begin{document}

\maketitle

\begin{abstract}This document is a companion for the Maple program
\textbf{Summing  a polynomial function   over integral points of a
polygon}.  It contains two parts. First, we see what this programs
does. In the second part, we briefly recall the mathematical
background.
\end{abstract}

\section{Introduction}

The present article is a user's guide for the Maple program
\textbf{Summing  a polynomial function over integral points of a
polygon}, available at
\verb"http://www.math.polytechnique.fr/~berline/maple.html". The
Maple program contains two types of computation. The first
computation does just what the title says. The input consists of a
finite set of rational points in $\Q^2$, whose convex hull is a
polygon $\p$, and a polynomial $h(x,y)$ with rational coefficients.
The output is the sum
$$
\sum_{(x,y)\in\p\cap\Z^2}h(x,y).
$$
The second computation returns the function of $t\in\N$ which arises
when the polytope $\p$ is dilated by $t$.
$$
E(t):=\sum_{(x,y)\in t\p\cap\Z^2}h(x,y).
$$
This function is a \emph{quasi-polynomial}, meaning that is has the
form
$$
E(t)=\sum_{i=0}^{\deg h+2}E_i(t)\; t^i,
$$
where the coefficients depend only on $t$ mod $q$, where $q$ is the
smallest integer such that $q\p$ has integral vertices. The function
$E(t)$  is called the  \emph{weighted Ehrhart quasi-polynomial} of
$\p$ with respect to the weight $h(x,y)$.

We apply two methods, the first one for a fixed polygon, the second
one for the computation of the weighted Ehrhart quasi-polynomial.
The first method is based directly on Brion's formula (\ref{brion}),
\cite{brion}, while the second method is based on the local
Euler-Maclaurin formula of \cite{EML}. Both methods use Barvinok's
decomposition into unimodular cones \cite{barvinok}. Although they
are very similar, the first method is faster when we deal with a
fixed polygon, while the second is faster when we want the Ehrhart
quasi-polynomial.

The software libraries \emph{LattE} \cite{latte} (improved version
in \cite{lattemacchiato}) and \emph{Barvinok} \cite{verdoolaege}
include the computation of the number of points of a rational
polytope in any dimension, together with many other applications.
Moreover, the weighted Ehrhart polynomials in any dimension are
computed in \emph{Barvinok}. The present program, in dimension two,
is  based on the same principles: Brion's formula and Barvinok's
decomposition of cones. We use however some new ideas on
"renormalisation" of Laurent series from \cite {EML} to speed up the
computation.  In the future, we will generalize it to higher
dimensions.

\section{Main commands}

\subsection{Summing a polynomial function over the set of  integral points of a
polygon} Let $P\subset \Q^2$ be a finite set of points.   Let
$\p\subset \R^2$ be the polygon obtained as the convex hull of the
set $P$. The program computes the sum
$$\sum_{(x,y)\in \p\cap \Z^2} h(x,y)$$ of the values of a polynomial $h(x,y)$ over
the set of integral  points contained in  $\p$. In particular, when
 $h=1$, it  computes the number of
integral points in $\p$.

 For a single monomial  $h(x,y)=x^{m_1}y^{m_2}$,
the command is
\begin{verbatim}
>sum_monomial_polygon(P,m);
\end{verbatim}
Here $P$ is a set of  pairs of rational numbers, and $m=[m_1,m_2]$
is a pair of non negative integers, the multidegree of the monomial
$x^{m_1}y^{m_2}$.

If we want just the number of integral of integral points, we can
use  the command
\begin{verbatim}
>number_points_polygon(P);
\end{verbatim}
This number  can be also obtained by the command
\begin{verbatim}
>sum_monomial_polygon(polygon,[0,0]);
\end{verbatim}
We  compute the sum of a polynomial $h(x,y)$ by  the command
\begin{verbatim}
>sum_polynomial_polygon(P,h);
\end{verbatim}
Here $P$ is a set of  pairs of rational numbers, and $h=\sum_m h_m
x^{m_1}y^{m_2}$ is entered as an expression in $x,y$.

\medskip

\noindent\textbf{Example 1.} $P$ is the square
$\{[0,0],[1,0],[1,1],[0,1]\}$.
\begin{verbatim}
>square:= {[0,0],[1,0],[1,1],[0,1]};

>number_points_polygon(square);
                               4
\end{verbatim}
The sum of values $x^5 y^5$ over the $4$ integral points  in the square is
\begin{verbatim}
>sum_monomial_polygon(square,[5,5 ]);
                             1\end{verbatim}
\noindent\textbf{Example 2.} Here $P$ is a randomly chosen set of 15 points.
\begin{verbatim}
> P := {[77/8,97/59], [93/44,70/29], [0,25/12], [25/32,29/48],
[92/41,57/91], [9/4,1/7], [64/43,31/75], [91/17,33/86], [12/37,77/8],
[8/5,41/27], [80/67,11/9], [16/73,11/89], [41/20,43/88],
[32/49,59/23], [77/94,65/46]}\end{verbatim}
The number of integral  points in the convex hull is $45$.
\begin{verbatim}
>number_points_polygon(P);
                             45\end{verbatim}
The vertices of the convex hull $\p$ of $P$ (listed in
counter-clockwise order) are obtained with  the {command}:
\begin{verbatim}
>vertices_in_counter_clock_order:=proc(polygon)
\end{verbatim}
\begin{verbatim}
>vertices_in_counter_clock_order(P);
 [[0,25/12], [16/73,11/89], [9/4,1/7], [91/17,33/86], [77/8,97/59],
[12/37,77/8]]
\end{verbatim}
We compute the sum of $x^{32}y^{32}$ over the set of  integral
points $(x,y)$  of the convex hull $\p$ of  $P$.
\begin{verbatim}
>sum_monomial_polygon(P,[32,32]);
          987532646688766560932727042325214847653263886
\end{verbatim}
We  compute the sum of $x^{32}y^{32}+7$ over all  the  integral
points $(x,y)$  of the polygon $\p$. (the preceding number $+7$
times $45$)
\begin{verbatim}
>h:= x^{32}y^{32}+7;
>sum_polynomial_polygon(P,h);
         987532646688766560932727042325214847653264201
\end{verbatim}

\subsection{ Weighted Ehrhart  polynomial of a polygon}

Our program computes also the weighted Ehrhart  quasi-polynomial of
a polygon. For brevity, we treat only the case where the weight is a
monomial $h(x,y)= x^{m_1} y^{m_2}$. When the polygon is dilated by a
non negative  integer $t$, and if $q$ is a positive integer such
that $q\p$ has integral vertices, the function of $t$ given by
$$t\mapsto
\sum_{(x,y)\in t\p\cap \Z^2} x^{m_1}y^{m_2}
$$
 is a quasi-polynomial $S(t)=\sum_{i=0}^{m_1+m_2+2} E_i(t)t^i$ of degree $m_1+m_2+2$.
The coefficients $E_i(t)$  are  functions of $t$ modulo $q$. This
program computes  these coefficients  $E_i(t)$  in terms of the
symbolic function $fmod(p*t,q)$ which stands for ($t \mapsto pt$ mod
$q$). We can either obtain each individual coefficient $E_i(t)$ or
the full weighted Ehrhart polynomial $S(t)$.

Here are the  commands:
\begin{verbatim}
> coeff_t_Ehrhart_polygon(i,t,P,m);
\end{verbatim}
 The input consists of   $i$ an integer, $t$ a letter , $P$ a set of points  and
 $m=[m_1,m_2]$ a pair of integers
which represents the weight; the output is the coefficient $E_i(t)$.

\begin{verbatim}
>Ehrhart_polynomial_polygon(t,P,m);
\end{verbatim}
Input is as in the previous command, except $i$ is not needed. The
output is the full Ehrhart polynomial $S(t)$.

\medskip
\noindent\textbf{Examples }
\begin{verbatim}
>transsquare:={[-1/2,-1/2]{[1/2,-1/2],[1/2,1/2],[-1/2,1/2]};
\end{verbatim}
\begin{verbatim}
> coeff_t_Ehrhart_polygon(0,t,square,[0,0]);
          1
\end{verbatim}
\begin{verbatim}
> coeff_t_Ehrhart_polygon(0,t,transsquare,[0,0]);
-2*fmod(t, 2)+3/2+1/2*fmod(t, 2)^2
\end{verbatim}
\begin{verbatim}
> Ehrhart_polynomial_polygon(t,square,[0,0]);
                          1 + 2 t + t^2
\end{verbatim}
\begin{verbatim}
> Ehrhart_polynomial_polygon(t,transsquare,[0,0]);
       (fmod(t, 2)-1)^2+(-2*fmod(t, 2)+2)*t+t^2
\end{verbatim}

\subsection{ Experiments}
The following experiments were done with a laptop, processor 1,86
GHz, RAM 782 MHz,0,99 Go.

\noindent\begin{verbatim}
>A:={[(-567337)/102495,-1414975/95662],[1/3,1/5],[-88141/20499,12732/47831]};

>largeA:={[1000*(-567337)/102495,1000*(-1414975/95662)],
[1000*1/3,1000*1/5],[-1000*88141/20499,1000*12732/47831]};

 > number_points_polygon(A);
                               36

> number_points_polygon(largeA);
                34922612
\end{verbatim}
In the next experiments, we indicate the time of computation $T$ in
seconds. The number of integral points in the rational triangle with
vertices  $A$ is $36$. If we dilate $A$ by the factor $1000$,  we
obtain the triangle $large A$ where the  number of points (34922612)
is approximatively $10^6$ times larger. Observe that we compute in
14 seconds the sum of the large degree monomial
$h(x,y)=x^{64}y^{64}$ over the set of integral points of $A$, and
that we compute in 16 seconds the sum of the same monomial over the
integral points of $largeA$. The computation time is almost the
same, although any computation by enumeration would be $10^6$ times
longer.
\begin{verbatim}
> T:=time(): sum_monomial_polygon(A,[32,32]);Time:=time()-T;
  11156693714080121436809683716369682546812787494001398139657
                           Time := 1.766

>  T:=time():sum_monomial_polygon(A,[64,64]);Time:=time()-T;
10691662746975383171690687952963005219723639375189814
217756070191566530558879\
  3836513555847334896253718879462978590217
                          Time := 13.640

> T:=time(): sum_monomial_polygon(largeA,[64,64]): Time:=time()-T;
17831035913722066043589677840496661987989193563450057671832979767102708226068\
  905195223428957659882216123374803724362290728944933635792703052976782671238\
  401601191375977184037799597789861617132380131198911864015293592136365221852\
  952449214916133197928922419462989960495593699297670700652853834584439172901\
  857916119694620105996329573478014513449383738873972550889051937620201341771\
  829110756841837358870588454172079624770000592845281113102517836579429050870\
  20099703621578931359063825440122383120351301766010118556183
                                Time:= 15.516


\end{verbatim}
Finally, we computed the weighted Ehrhart polynomial with weight
$x^{32}y^{32}$ over the triangle  with vertices[[-567337/102495,
-1414975/95662], [88141, 292844676/6833], [-88141/20499,
12732/47831]].  We compute the coefficient of $t^2$ for example. The
time of computation is
 268 seconds.  The result is too big to be printed here,
 as it involves many functions $fmod(c*t,D)$ where $D$ runs though the denominators
 of the coordinates of the vertices of $A$ (large numbers).
\begin{verbatim}
> T:=time():coeff_t_Ehrhart_polygon(2,t,[[-567337/102495,
-1414975/95662], [88141, 292844676/6833], [-88141/20499,
12732/47831]],[32,32]): Time:=time()-T;
                          Time :=  268.266
\end{verbatim}

\section{ Mathematical background}

The first method is  for a fixed polygon, the second one for the
computation of the weighted Ehrhart quasi-polynomial.

\subsection{ First method: Brion's formula, Barvinok's decomposition
into unimodular cones  and iterated Laurent series}

Let $\p$ be a convex polygon in $\R^2$ with rational
vertices $s_i,1\leq i\leq n+1$.  We want to compute the sum
\begin{equation}\label{somme}
\sum_{x\in \p\cap \Z^2} x^{m_1} y^{m_2}.
\end{equation}
We start by observing that (\ref{somme}) is equal to the
coefficient of $\frac{\xi_1^{m_1}\xi_2^{m_2}}{m_1! m_2!}$ in
$$
\sum_{x\in \p\cap \Z^2} e^{\langle \xi,x\rangle}.
$$
 Our method is based on Brion's formula (\ref{brion}).
Brion's formula  is the generalization of the following formula for
the  sum of geometric progressions over the interval $[A,B]$ (with
$A\leq B$ integers):
$$\sum_{A}^B e^{n \xi}= \frac{e^{A\xi}}{1-e^{\xi}}+ \frac{e^{B\xi}}{1-e^{-\xi}}.$$
For any rational polygon  $\q\subset\R^2$ define
$$
S(\q)(\xi)=\sum_{x\in \q\cap \Z^2} e^{\langle \xi,x\rangle}.
$$
This a meromorphic function near $\xi=0$.
Moreover the map $\q\mapsto S(\q)(\xi)$ is a valuation on the set of rational polyhedra, and
$S(\q)=0$ if  $\q$  contains a line.
Brion's formula is the following.
Let $\c_i$ be the cone at vertex $S_i$ of the polygon.
\begin{equation}\label{brion}
S(\p) =\sum_{i=1}^{n+1}S(\c_i).
\end{equation}
Each term $S(\c_i)(\xi)=\sum_{x\in \c_i \cap \Z^2} e^{\langle
\xi,x\rangle}$ in (\ref{brion}) is a meromorphic function near
$\xi=0$. The poles cancel and the sum is a holomorphic function of
$\xi$. Thus we compute (\ref{somme}) as the coefficient of
$\frac{\xi_1^{m_1}\xi_2^{m_2}}{m_1! m_2!}$  in the right-hand-side
of (\ref{brion}). We actually  compute  the individual
 contribution of each  cone $\c_i$ (associated to the vertex $s_i$)  to the sum.
The coefficient of $\frac{\xi_1^{m_1}\xi_2^{m_2}}{m_1! m_2!}$  in
the meromorphic function $S(\c_i)$ of two variables $\xi_1,\xi_2$
has no intrinsic meaning. Our method  consists in applying {\bf
iterated Laurent series} expansions to $S(\c_i)(\xi)$  with respect
to the variables $\xi_1$ then $\xi_2$. We obtain a Laurent series
$L(\c_i)$  in the ring $\Q[\xi_1,\xi_1^{-1},\xi_2,\xi_2^{-1}]$ and
we compute  the coefficient $\frac{\xi_1^{m_1}\xi_2^{m_2}}{m_1!
m_2!}$  in  $L(\c_i)$.

\medskip

Thus, in order to compute the contribution of a vertex $s$ to the sum
(\ref{brion}), we need to compute $S(\c )( \xi)$ for the
 supporting cone  $\c$.   The crucial tool here is
Barvinok's decomposition into unimodular cones.
Actually, we use the following variant of Barvinok's decomposition,
(procedure \protect\verb+signed_decomp+).

Let $\c$ be  a simplicial cone in $\R^d$. Let  $V_i$, for
$i=1,\ldots, d$, be the generators of $\c$. Let $V$ be a vector in
$\R^d$. We write $V=\sum_{i}u_i V_i$. We split $[V_1,\dots,V_d]$
into three parts, as follows.
$$
L_+:=[X_1,\ldots, X_k]
$$
formed by the $V_i$
 such that $u_i>0$,
$$
L_-:=[Y_1,\ldots, Y_m]
$$
formed by the $V_i$
 such that $u_i<0$,
$$
L_0:=\{Z_1,\ldots, Z_b\}
$$
formed by the $V_i$  such that $u_i=0$.

Then we have the equality of characteristic functions modulo
characteristic functions of cones containing lines.
\begin{eqnarray*}
 && (-1)^{(k+1)}[\c]=\sum_{i=1}^k (-1)^{i+1}
[\c(X_1,\ldots,X_{i-1},-X_{i+1},\ldots,-X_k,V,L_{-},L_0)]+\\
&&\sum_{j=1}^{m}(-1)^{j+k}
[\c(L_+,-V,-Y_1,\ldots,-Y_{j-1},Y_{j+1},\ldots,Y_m,L_0)].
\end{eqnarray*}
\noindent\textbf{Remark}. This decomposition is not the stellar
decomposition. It involves only cones of maximal dimension $d$. It
avoids the dualizing trick of Brion.

 \noindent\textbf{Example}.
$\c=\R^+ e_1\oplus \R^+ e_2$, $V=e_1+e_2$, so that $L_-$ and $L_0$
are empty and $k=2$. Then
$$
-[\c]=\c(V,-e_2)-\c(e_1,V)-\c[e_2,-e_2,e_1].
$$
Indeed $[\c(e_2,-e_2,e_1)]-[\c]$ is equal to the characteristic
function of the quadrant $(e_1, -e_2)$ minus that of the  half-line $\R^+ e_1$.
This is also the case for $[\c(V,-e_2)]-[\c(e_1,V)]$.

\medskip

 If we use a lattice vector $V$
with sufficiently small  coordinates in the basis $(V_i)$, the cones
appearing in this decomposition have  indices smaller than $\c$. One
obtains such a \emph{short} vector $V$ by the Lenstra-Lenstra-Lovasz
algorithm. By a repeated application of this decomposition, one
obtains a decomposition of $\c$ in a signed sum of unimodular cones
$\c_z$ (modulo cones containing lines). As $S({\mathfrak a})=0$ for
a cone   ${\mathfrak a}$ which contains a line, we can  use this
decomposition to compute $S(\c)$.

For a unimodular cone $\c$, the sum $S(\c)$ has a simple closed
expression. Let $(V_1,V_2)$ be primitive generators of the edges of
$\c$ and let $s$ be its vertex. Let $\tilde{s}$ be the unique
integral point contained in the \emph{semi-closed box}
$$
\{s+t_1 V_1+t_2 V_2,  0\leq t_i <1\}
$$
If $s=s_1 V_1+s_2 V_2$, then $\tilde{s}=\tilde{s}_1 V_1+\tilde{s}_2
V_2$ with $\tilde{s}_i=ceil(s_i)$. Then
\begin{equation}\label{unimodular}
S(\c)(\xi)=\frac{e^{\langle \xi,\tilde{s}\rangle}}{(1-e^{\langle \xi,V_1\rangle})(1-e^{\langle \xi,V_2\rangle})}.
\end{equation}

\noindent In order to simplify the computation of iterated Laurent
series, we introduce the analytic function
$$B(X,u)=\frac{e^{u X}}{1-e^X}+\frac{1}{X}=
-\sum_{n=0}^\infty \frac{b(n+1,u)}{(n+1)!} X^n $$ where $b(n,u)$ are
the Bernoulli polynomials. Writing
\begin{equation}\label{bernouilli}
\frac{e^{u X}}{1-e^X}= B(X,u)-\frac{1}{X},
\end{equation}
we obtain
$$
S(\c)(\xi)=A+G+R,
$$
where  $$A=B(\langle \xi,V_1\rangle, \tilde{s}_1) B(\langle
\xi,V_2\rangle, \tilde{s}_2)$$ is an analytic function of $\xi$,
 $$G:=-\frac{1}{\langle \xi,V_1\rangle} B( \langle \xi,V_2\rangle, \tilde{s}_2)-
\frac{1}{\langle \xi,V_2\rangle}B( \langle \xi,V_1\rangle,
\tilde{s}_1),
$$
$$R:=\frac{1}{\langle \xi,V_1\rangle \langle \xi,V_2\rangle}.
$$
We replace  $\frac{1}{\langle \xi,V_1\rangle}$ and  $\frac{1}{\langle \xi,V_2\rangle}$  by their iterated Laurent series expansion
in the ring $R[\xi_1, \xi_1^{-1}, \xi_2,\xi_2^{-1}].$
For example, if $V_1=[2,1]$, we write
$$\frac{1}{2\xi_1+\xi_2}=\frac{1}{\xi_2}\frac{1}{(1+2\xi_1/\xi_2)}=\frac{1}{\xi_2}\sum_{k=0}^{\infty} (-1)^k
2^k (\xi_1/\xi_2)^k.
$$
We then replace  $S(\c)$  by  the corresponding  element in
$\Q[[\xi_1,\xi_2,\xi^{-1},\xi^{-2}]]$ and we take the coefficient of
$\xi_1^{m_1}\xi_2^{m_2}$.

\bigskip \noindent {\bf Remark } The weighted Ehrhart polynomial can also be computed  by this method. We
did not write the corresponding algorithm in the Maple file, because
we observed that a faster algorithm is given by the second method
which we describe in the next section. However, let us explain what
one should do. When the polytope $\p$ is dilated in $t\p$,    its
vertices are dilated by $t$, while the edges of the cones at
vertices do not change. Thus we have to compute
 \begin{equation}\label{unimodularpar}
S(\c_t)(\xi)=\frac{e^{\langle \xi,\tilde{s_t}\rangle}}{(1-e^{\langle \xi,V_1\rangle})(1-e^{\langle \xi,V_2\rangle})}
\end{equation}
where now $s_t$ is the  unique point  with integral coordinates in the box
$$
\{ts+u_1 V_1+u_2 V_2,  0\leq u_i <1\}.
$$
 If $s=s_1 V_1+s_2 V_2$, with $s_i=p_i/q_i$,
  we see that $$s_t= [t s_1+mod(-tp_1,q_1)/q_1,
  ts_2-mod(-tp_2,q_2)/q_2].$$
The iterated Laurent series  in $\xi_1,\xi_2$ has  coefficients which are
  polynomials in $t$ and the periodic  functions  $mod(tp_i,q_i)$.
 We extract the coefficient of  $t^j \xi_1^{m_1}\xi_2^{m_2}$.
\subsection{ Second method. Weighted  Ehrhart
quasi-polynomial using local Euler-Maclaurin formula}
  We now recall the results of \cite{EML}  and explain how they  can be applied  to the
computation of the weighted Ehrhart quasi-polynomials. Let $\p$ be a
convex polytope in   $\R^d$, with rational vertices. Let $h(x)$ be a
polynomial function of degree $r$ on $\R^d$. We want to compute the
sum $\sum_{x\in \p\cap\lattice}h(x)$  of values $h(x)$ over the set
of integral points of the polytope $\p$.

The local Euler-Maclaurin formula has the following form.
\begin{equation}\label{maclaurin}
\sum_{x\in \p\cap\lattice}h(x)=\sum_{\f\in\CF(\p)}\int_\f
D(\p,\f)\cdot h
\end{equation}
where $\CF(\p)$ is the set of all faces of $\p$. For each face $\f$,
$D(\p,\f)$ is a   differential operator of infinite degree with
constant coefficients associated to $\f$. The operator $D(\p,\f)$ is
\emph{local}, in the sense that it depends only on the intersection
of $\p$ with a neighborhood of any generic point of $\f$. The
integral on the face $\f$ is taken with respect to the Lebesgue
measure on $<\f>$ defined by the lattice $\Z^d\cap \lin(\f)$.  Here
$<\f>$ is  the affine span of the face $\f$ and $\lin(\f)$ is the
linear subspace parallel to $<\f>$.

Let us recall the construction of the operators $D(\p,\f)$. We
denote by $\t(\p,\f)$ the transverse cone to $\p$ along $\f$. Using
the standard scalar product, $\t(\p,\f)$ is described as the
following affine cone in $\R^d$. Let $\lin(\f)^\perp$ be the vector
subspace orthogonal to $\lin(\f)$. Then $\t(\p,\f)$ is the
orthogonal projection  on $\lin(\f)^\perp$ of the supporting cone of
$\p$ along $\f$. The operator $D(\p,\f)$ is defined in terms of the
transverse cone $\t(\p,\f)$, as follows.

For every rational affine cone $\a\subset V$, we construct in
\cite{EML} an analytic function $\xi\mapsto \mu(\a)(\xi)$ on $\R^d$.
This construction depends on the choice of a scalar product. Here we
use the standard scalar product. These functions $\mu(\a)$ have nice
properties which play a crucial role in our method. First, the
assigment  $\a\mapsto\mu(\a )$ is a \emph{valuation} on the set of
affine cones with a given vertex. Second, it is \emph{invariant
under lattice translations.}
 Furthermore,  $\mu(\a)=0$ \emph{if}  $\a$ {\em contains a line.}

We define
$$ D(\p,\f)=  D(\mu(\t(\p,\f)))
$$
as the differential operator of infinite degree with constant
coefficients, with symbol $\mu(\t(\p,\f))(\xi)$. In other words, if
$\xi=(\xi_1,\dots ,\xi_d)$, we obtain $ D(\p,\f)$ by replacing
$\xi_i$ by $\frac{\partial}{\partial x_{i}}$ in the Taylor series of
$\mu(\t(\p,\f))(\xi)$.

 For
any positive integer $t$, we consider the dilated polytope $t\p$ and
the corresponding sum
$$
S(t\p,h)= \sum_{x\in t\p\cap\lattice}h(x).
$$
 From
(\ref{maclaurin}), it follows easily that the function  $t\mapsto
S(t\p,h)$ is given by a quasi-polynomial: there exist periodic
functions $t\mapsto E_i(\p,h,t)$ on $\N$  such that
\begin{equation}\label{coeffEhrhart}
S(t\p,h)= \sum_{i=0}^{d+r} E_i(\p,h,t)t ^i
\end{equation}
whenever $t$ is a positive integer. Moreover the coefficients
$E_i(\p,h,t)$ are computed using the functions $\mu(\t(t\p,t\f))$.
 Indeed, let $s$ be the vertex of
$\t(\p,\f)$ so that
 $\t(\p,\f)=s +\t_0$. Then the dilated transverse cone is $\t(t\p,t\f) = t s + \t_0$.
As $\a\mapsto\mu(\a )$ is invariant under lattice translations, we
have
$$
\mu(\t((t+q)\p,t\f))=\mu(\t(t\p,t\f)),
$$
 if $q$ is an
integer such that $q s$ is a lattice point for the projected
lattice, or equivalently, such that $q<\f>$ contains a lattice
point. Thus, the coefficients $E_i(\p,h,t)$ depend only on $t\mod
q$, where $q$ is  the smallest integer such that $q\p$ has integral
vertices.

When $\a$ is a \emph{unimodular} affine cone of dimension $1$ or
$2$, the functions $\mu(\a)$ have an explicit form, in terms of the
functions $B(X,u)$ introduced in (\ref{bernouilli}).

 Let $\d$ be  a one dimensional affine
cone of the form $(s + \R_+)V$ where $V$ is a primitive vector and
$s\in\Q$. We have
\begin{equation}\label{mudim1}
\mu(\d)(\xi)= B(\langle \xi,V \rangle,ceil(s)-s).
\end{equation}
Let $\a$ be a two dimensional  \emph{unimodular} affine cone. Let
$V_1,V_2$ be primitive generators of its edges, such that
$\det(V_1,V_2)=1$. For  $\xi=(\xi_1,\xi_2)\in \R^2$, let
$y_i=\langle\xi,V_i\rangle$ , for $i=1,2$, be the coordinates of
$\xi$ relative to the dual basis $(V_1^*,V_2^*)$. We write the
vertex of $\a$ as $s_1 V_1 + s_2 V_2$ with $s_i\in\Q$. Let
$\epsilon_i= ceil(s_i)-s_i$, and let $C_i= \frac{\langle
V_1,V_2\rangle }{\langle V_i,V_i\rangle}$, for $i=1,2$. With these
notations, we have
\begin{multline}\label{unimodulaire}
\mu(\a)(\xi)= \frac{e^{\epsilon_1 y_1+\epsilon_2 y_2}} {
(1-e^{y_1})(1-e^{y_2})} +\\
\frac{1}{y_1} B(y_2-C_1y_1,\epsilon_2) + \frac{1}{y_2 } B(y_1- C_2
y_2,\epsilon_1)- \frac{1}{y_1 y_2}.
\end{multline}
 The function
$\mu(\a)(\xi)$ is actually analytic, although this  is not obvious
on (\ref{unimodulaire}). In order to compute the contribution of a
vertex $s$ of $\p$ to the sum (\ref{maclaurin}), we need to compute
$\mu(\c)(\xi) $ when $\c$ is  the two-dimensional supporting cone at
$s $. The crucial tool here is Barvinok's decomposition into
unimodular cones.  The valuation property of $\a\mapsto\mu(\a )$
makes it possible to reduce the computation to the unimodular case,
and use (\ref{unimodulaire}). Notice that (\ref{unimodulaire})
returns a function of the relative coordinates $(y_1,y_2)$, which we
must convert back to a function of the standard coordinates
$(\xi_1,\xi_2)$, in order to add the contributions of the various
unimodular cones in Barvinok's decomposition. Actually, since
$\mu(\a)=0$ if the cone $\a$ contains a line, we use the variant of
Barvinok's decomposition described in the first method.

\end{document}